# Self-Organized Criticality: A Guide to Water-Protein Landscape Evolution

J. C. Phillips

Dept. of Physics and Astronomy, Rutgers University, Piscataway, N. J., 08854

Abstract

We focus here on the scaling properties of small interspecies differences between red cone opsin transmembrane proteins, using a hydropathic elastic roughening tool previously applied to the rhodopsin rod transmembrane proteins. This tool is based on a non-Euclidean hydropathic metric realistically rooted in the atomic coordinates of 5526 protein segments, which thereby encapsulates universal non-Euclidean long-range differential geometrical features of water films enveloping globular proteins in the Protein Data Bank. Whereas the rhodopsin blue rod water films are smoothest in humans, the red cone opsins' water films are optimized in cats and elephants, consistent with protein species landscapes that evolve differently in different contexts. We also analyze red cone opsins in the chromatophore-containing family of chameleons, snakes, zebrafish and goldfish, where short- and long-range (BLAST and hydropathic) aa correlations are found with values as large as 97-99%. We use hydropathic amino acid (aa) optimization to estimate the maximum number $N_{max}$ of color shades that the human eye can discriminate, and obtain $10^6 < N_{max} < 10^7$, in good agreement with experiment.

Many complex systems are optimized, or nearly so, but proteins are the most extensively optimized and best-documented such systems. Because proteins are so complex - there are $10^{385}$ possible aa sequences for a typical transmembrane protein with 350 aa side groups on its peptide chain – it has long been assumed that the only logical approach to understanding their combined functional and species dependencies lay through amassing extensive data of all kinds: amino acid sequences, genomes, crystal structures, and many properties. It appeared that there were no unifying "short cuts" through this exponentially complex maze, and this assumption seemed to be borne out by the limited nature of the results obtained from energy landscapes generated by a wide variety of Newtonian simulations of small protein folding[1]. Here we will show that a landscape approach based not on energy but on the scaling of long-range differential geometry of water-protein interfaces reveals new aspects of the complex function-species maze, and yields remarkable results even at the cellular length scale (about $10^2$ longer than the molecular scale).

The basis of our approach is the discovery[2] that the evolution of solvent accessible surface areas (SASA, for a 2 A water molecule) of protein segments of length 2N+1 exhibit power-law scaling (Self-Organized Criticality, SOC) at long range with $4 \leq N \leq 17$, described by the long-range SOC critical exponents $\Psi = \{\psi(aa)\}$. These are defined by the MZ dimensionless exponents

$$\text{dlog(SASA(aa))/dlogN} = -\psi(aa) \qquad (1)$$

Protein functionality is often determined by teretiary conformational changes guided by weak, long-range hydropathic interactions ($N \geq 4$), rather than strong, short-range steric packing secondary (α helical and β strand) ones ($N < 4$), and this is why SOC can be a magic wand for protein physics[3].

Previously we introduced the concept of hierarchical hydropathic roughening of water-membrane protein interfaces[3], as reflected in the variances $R(W)$ and $R^*(W)$ of sliding window profiles of $\psi(aa)$ averaged over windows of length W. Here W is not merely a convenient parameter which functions as a tunable length scale. By correlating it with evolutionary trends, we can identify the length scale that is relevant for optimizing protein functionality either *ex situ* ($R(W)$) or *in situ* ($R^*(W)$). For large W (appropriate to membrane proteins, where the membrane thickness W ~ 25), we compared these alphanumerical roughness variances $R^*(W)$ to the alphabetical aa contact similarities defined by BLAST, the standard Web-based multiple alignment tool, denoted here by $B(P_i,P_j) = B_{ij}$. We studied rhodopsin, the night vision transmembrane rod receptor associated with night vision in species ranging from mammals to bacteria. Much to our surprise, we established a remarkably accurate (as high as 96% for W = 25) correlation between $R^*(W)$ and $B(human,P_j)$, for $P_j$ = five species (humans, monkeys, cats, mice and rabbits). Here we investigate similarly successful correlations for the red cone receptor, and use them to illustrate the comparative advantages of the hierarchical alphanumerical hydropathicity scale. At the same time, our correlations show that BLAST contains various kinds of previously hidden information that are established through the correlations with $R^*(W)$. This hidden information goes well beyond the sequence similarities provided directly by BLAST, and yields new insights into evolution of properties and sequential trends.

There are several technical points to consider. The hierarchical order of $B(P_i,P_j)$ for fixed $P_i$ as a function of $P_j$ depends on the choice of reference protein $P_i$. This is analogous to a well-known problem in field theory, where many bilinear algebraic models are gauge-dependent, as contrasted with Maxwell's equations for electromagnetic fields, which are gauge-invariant (independent of an additive constant in the vector potential, here analogous to the choice of reference protein). This gauge dependence can actually be used to advantage in analyzing certain large-scale problems, for instance, the protein dependence of the vertebrate – invertebrate species separation, which will be discussed elsewhere. Here we acknowledge its existence and utilize it to clarify the different roles played by rod and cone opsins in retinal mosaics and ganglion networks.

The molecular structures of rod and cone photoreceptors are different and vary substantially from species to species, even for the proximate subfamily of humans, monkeys, cats, mice and rabbits[4]. Although outnumbered more than 20:1 by peripheral rod photoreceptors, cone cells, which are sharply clustered and centered as largely hexagonal mosaic arrays[5] in the human retina, mediate daylight vision and are critical for visual acuity and color discrimination. Both striking similarities and well-resolved differences between red opsin cone and blue rhodopsin rod aa sequences and hierarchical ordering of $B(P_i,P_j)$ for fixed $P_i$ as a function of $P_j$, as well as excellent correlations with $R^*(W)$, are found below for this mammalian subfamily.

Our first step was to compare human red cone roughness profiles $R^*(W)$ with those of rhodopsin over a wide range of species, including invertebrates which have compound eyes (see Fig. 1; because of their complexity, the data are presented graphically, rather than as tables). This comparison shows a much larger hydropathic roughness for human red cone than other species' rhodopsin rod, which begins around length scale $W = 9$, which is just where Eqn. (1) becomes valid, and the long-range hydropathic interactions become stronger than the short-range steric packing interactions[2]. The figure also shows several other interesting evolutionary features, including the fact that for $W < 25$ (membrane localization), the smoothest rhodopsin species are invertebrates (multilens ant and fruit fly), while for $W > 25$ rhodopsin $R^*(47)$ for invertebrates become as rough as red cone opsin human.

Differences in retinal morphology explain the qualitative differences between vertebrate and invertebrate rhodopsin roughness profiles $R^*(W)$. The small-scale cellular structure of compound eyes enables insects to utilize smoother rhodopsin, which probably enhances their visual temporal resolution (200 images/sec in bees, compared to 30 images/sec in humans). Invertebrates have compound eyes with 2 primary cells and ~ 10 secondary pigment cells[6] per retina, compared to ~ $10^5$ rod cells/retina in mammals[4]. This difference corresponds to a change in lateral molecular correlation lengths of a factor of 100. The smoother invertebrate roughness profiles $R^*(W)$ for $W < 25$ reflect a trade-off between long-range smoothness for $W > 25$, which is ineffective with only 10 pigment cells/retina, in favor of smoother photoreceptor-membrane interactions in their ommatidia. It is striking that the retinal morphological differences (compound-single lens) between vertebrate and invertebrates are qualitatively obvious in their rhodopsin roughness profiles $R^*(W)$. In principle the dominance of $R^*(W)$ for $W > 25$ in large mammalian retina reflects the nearly scale-free property at large lengths that characterizes SOC in more evolved species.

Given these differences, can one explain the much larger hydropathic roughness for human red cone than rhodopsin rod? The polychromatic cone opsins are located at the center of the retinal area, and are supported there by the rod peripheral mosaic, with its 15x larger area. The small size of the cone area greatly reduces the length scale for the cones, and this contributes to increased chromatic cone roughening. A second factor is that the overall roughness of cone opsins increases with the maximum absorption peak wavelength (for human, long, 560 nm, medium, 530 nm, short, 420 nm). Site-directed mutagenesis studies for a large number of vertebrate M/LWS pigments correlate with these shifts (the "three-site rule")[7]. However, the same variation can also be explained, without the disruptive effects of mutagenesis, from hydropathic correlations with wild-type cone opsin $R^*(W)$ alone. Because there are only three variables, the correlations for $9 \leq W \leq 41$ are excellent, being 0.968 for W = 9 and 0.979 for W = 41. The striking result is that the best correlation occurs for W = 21, and it gives R = 0.988. If nearly the entire transmembrane region $W \leq 25$ participates in optimizing optical signaling, then one would expect the best correlation to occur around W = 21, as it does. This suggests that the global cone opsin aa sequence *in situ* roughness $R^*(W)$, which measures the long-range water-protein interactions, also is an important factor in determining maximum absorption peak wavelength shifts.

To begin the search for the best red cone opsin correlation of $B(P_i,P_j)$ for fixed $P_i$ as a function of $P_j$ with $R^*(W)$, we first construct the $R^*(W)$ profiles for candidate human proximate red opsins for which reviewed aa sequences are available from Uniprot. The candidate proteins were human, cat, goat, dog, elephant, bovine and monkey with the latter's aa sequence kindly supplied by Prof. S. Kawamura[8]. While rhodopsin rod bovine is quite smooth and stable[3], red cone bovine is much rougher than the other proximate $R^*(W)$, and it was discarded as a statistical outlier. The roughness profiles for the six remaining proteins are shown in Fig. 2: note that the interspecies differences are small (typically < 10%). Next we constructed the BLAST similarity matrix. For rhodopsin we found (Table II of ref. 3) that human roughness profiles $R^*(W)$ were smoothest for $3 \leq W \leq 47$, but for the red cone opsins we find (Fig. 2) a much narrower and less regular pattern, which is not surprising considering that the more loosely spaced cone opsins are concentrated in a small central area and surrounded by a large, nearly regular, more closely spaced almost crystalline rod mosaic[4,9].

There are two leading candidates for least roughening, elephants and cats, so we have calculated the correlation of B($P_i$,$P_j$) for fixed $P_i$ for these two cases, first for $P_i$ = elephant, and then for $P_i$ = cat. For $P_i$ = elephant as a function of $P_j$ the correlation with $R^*(W)$ peaks at 77% for W = 25. The $P_i$ = cat-based correlation reaches 88% at W = 25 and peaks at 90% for W = 47, so the cat correlation (used as baseline in Fig. 2) is clearly superior, in contrast to rhodopsin rods, where humans correlated best. The greater smoothness for large W of elephant red opsin compared to cat and other species' red opsin could be the result of the much larger elephant mass and elephant dimensions. Human red opsin is very similar to monkey red opsin, as expected, while goat and dog exhibit contrary difference profiles. This could reflect carnivore/herbivore differences. The larger smoothness for W ~ 50 may provide predatory dogs with faster visual signaling (faster signal transmission across membranes).

We have also analyzed red cone opsins in the chromatophore-containing family of chameleons, snakes, zebrafish and goldfish. The BLAST patterns show strong chameleon – snake and zebrafish - goldfish paired correlations, so conventional B - $R^*$ correlation coefficients could contain little information. Although details are not known, chromatophore functionality apparently relies on pigment redistribution due to motile activities of specialized cells, which occurs in nonmammalian vertebrates in a wide array of extraocular tissues, including the pineal complex and brain[10]. In mammals a wide range of non-canonical opsins has been identified[11], which may all be related to red cone retinal opsins through gene duplication.

These ideas can be tested hierarchically against $R^*(W)$ alone, without involving BLAST correlations. As shown in Fig. 3, $R^*(W)$ for the four species are all similar up to W = 21, but upon reaching W = 25 (TM L = 1), chameleon – snake and zebrafish - goldfish pair off as they should, with $R^*(W)$ increasing rapidly for chameleon – snake compared to zebrafish – goldfish. The intrapair hierarchical ordering also appears natural (chameleon smoother than snake, goldfish (solid color) smoother than multicolor zebrafish). These successes encourage a full search for optimized $B - R^*$ correlations (4 species), with the following results: for fixed $P_i$ = goldfish, 97% correlation for W = 47, while for fixed $P_i$ = chameleon, 99% correlation for W = 7. Given the remarkably sensitive protective responses of chameleon[12], this small value of its optimized W is not unreasonable, as it maximizes the coloric environmental replication.

Before leaving the subject of color vision, we estimate a fundamental quantity, the number of color shades that the human eye can discriminate. Human vision is trichromatic. Empirically it is known that color discrimination (or quantization) can be described by 8, 8 and 7 bytes/color channel, and that this discrimination is almost perfectly multiplicative, leading to a rough estimation of the number of distinguishable colors in the optimal color space as $3.2.10^6$ (22 bytes)[13]. Retinal organization in primates, which have a complex visual behavioral repertoire, appears relatively simple. However, quantitative connections between primate vision, retinal electronic connectivity (or neural network) and color quantization have so far surprisingly not emerged from studies of physiological models[14]. Here we argue that the primate vision system is nearly perfectly optimized by a simple rhodopsin amino acid (aa) optoelectronic signal (retinal upload)/(pigment download), and that this system is another canonical example of long-range chemical and mechanical self-organization of protein aa sequences implied by SOC.

The separation of visual signals into three channels will be most efficient if those channels are nearly equally informative, and this is found to be the case[15]. This implies that the primate 20 aa opsin sequence is nearly optimally designed to interact with all three pigment download paths, and that these paths can be primarily influenced by opsin aa hydrophobicities. The MZ hydrophobicity scale[2] is based on solvent-accessible surface area exponents ψ(aa) based on the self-organized criticality (SOC) of globular protein folding. The average spacing of these ψ(aa) exponents is 0.008, and each aa can have a distinctive dynamical contribution if its hydrophobicity is well separated from other aa by 0.004. Even if the MZ hydrophobicities are not well separated, two aa may still make distinctive contributions if their strongly polarizable ring contents are different. Combining these two criteria, optoelectronic hydroredundancy occurs in only three cases: for one-ring Tyrosine (ψ(Y) = 0.222) and Phenylalanine (ψ(F) = 0.218) and aliphatic chain (no rings) Isoleucine (ψ(I) = 0.222) and Methonine (ψ(M) = 0.221), and Alanine (ψ(A) =0.157) and Glycine (ψ(G) = 0.156). Thus there are effectively 17 remaining functionally independent aa available, with a byte content of 4.12. Nearly optimal two-step uploading and downloading can resolve 8.24 ~ 8 bytes/tricolor channel, in agreement with the empirical CIELAB color difference formula (8,8,7) quoted above[13].

In earlier work[16] we found that humans have the smoothest rhodopsin, which might suggest that the most evolved species always have the most evolved proteins. This anthropomorphic picture is contradicted here by our present result, which is that the smoothest red opsins are found in cats (nocturnal predators). Note also the elephant's smooth red opsin; elephants often search for food in deep shade with eyes far removed from areas searched.

The scaling properties of small interspecies differences discussed here are generally inaccessible to direct observation through structural studies[17]. They may be of great importance in analyzing intraspecies mutational differences in mutationally prolific viruses (for example, HIV). While the various examples discussed here are hopefully valuable and interesting in themselves, the unprecedented success of the SOC-based magic wand discussed here has larger implications. A common thread runs through all our studies of exponentially complex, self-organized networks, including such apparently unrelated systems as network glasses[18], ceramic high-temperature superconductors[19], the citation web of 20$^{th}$ century science (25 million papers, 600 million citations)[20,21,22], and proteins[2,3]. It appears that this common thread is mathematical and is based on modern non-Euclidean and non-Newtonian mathematical tools, such as differential geometry, set theory, and topology. The connection with differential geometry[23] is especially clear here in the context of the long-range evolution of solvent accessible surface areas[2,17] that appear in water film protein packaging.

**Materials and Methods**

The values of $\psi(aa)$ are listed in Table 1 of ref. 17. The specific Uniprot sequences in Fig. 1 are P08100, P04000, P31356, Q17292, P87369, Q9UHM6, O01668; in Fig. 2 are P04000, (GenBank: AB193772), O18913, Q95170, O18914, Q68J45, and in Fig. 3, Q9W6A7, P41592, P32313, C6G443.

**Figure Captions**

Fig. 1. Here the roughening form factor of human red cone opsin is compared to vertebrate and invertebrate rod rhodopsin form factors (all normalized against human rhodopsin, the "best" rhodopsin[3]). See text for discussion. Note the wide ordinate range. Here Mela refers to melanopsin[11].

Fig. 2. The roughening form factors of red mammalian cone opsins span a narrow range and suggest either elephant or cat as "best" (see text).

Fig. 3. The roughening form factors of red cone opsins in the chromatophore-containing family discussed in the text.

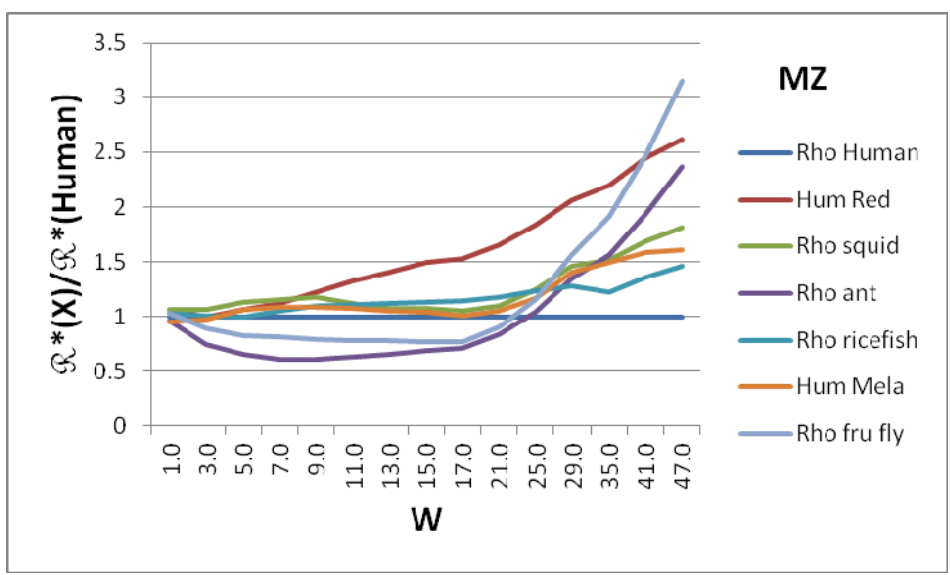

Fig. 1

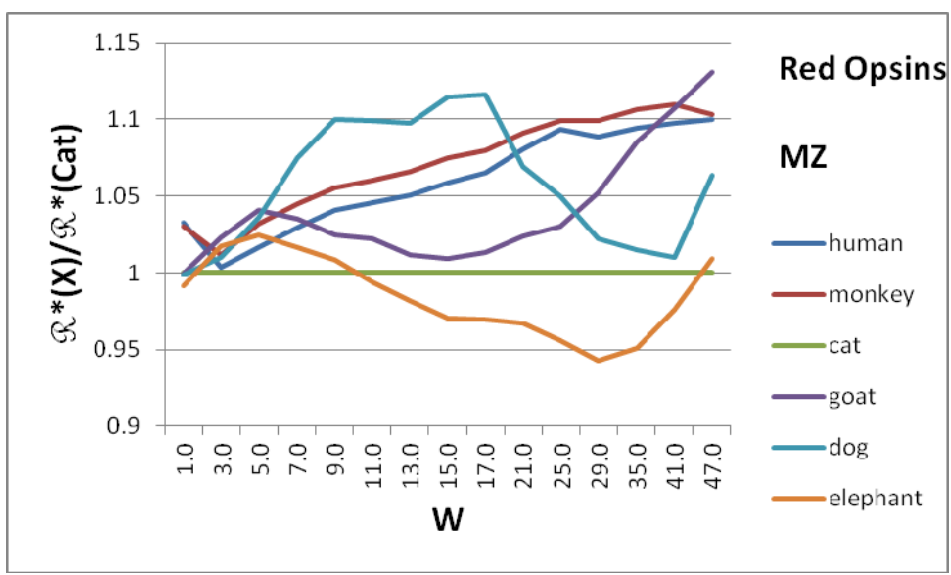

Fig. 2

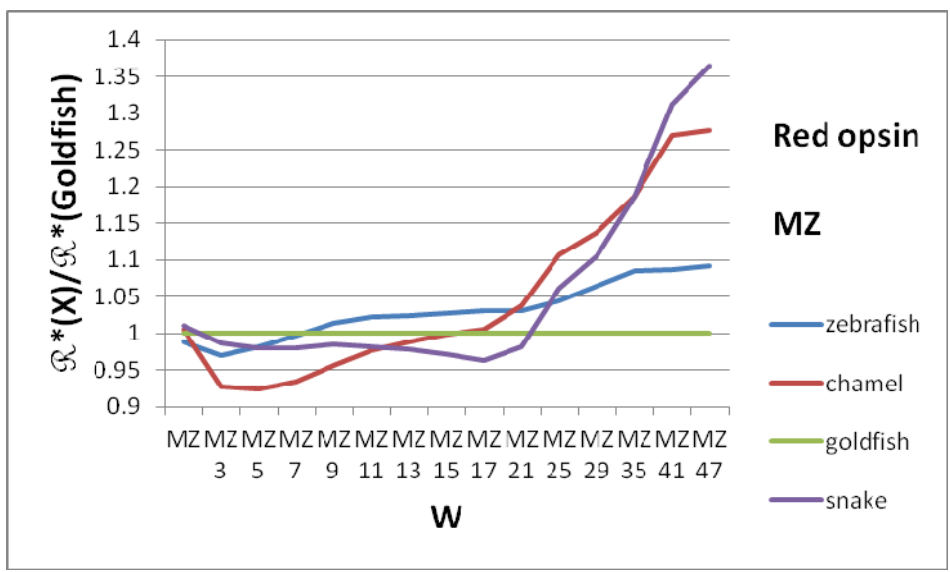

Fig. 3

16